\documentclass[traditabstract]{aa}

\usepackage[colorlinks=true, linkcolor=blue, citecolor=blue, urlcolor=blue]{hyperref}
\usepackage{amsmath}
\usepackage{graphicx}
\usepackage{txfonts}
\usepackage{amssymb}
\usepackage[]{natbib}
\usepackage{mathrsfs}
\usepackage{caption}
\usepackage{float}
\usepackage{afterpage}
\usepackage{amstext}

\begin{document}

\title{First Light of Engineered Diffusers at the Nordic Optical Telescope Reveal Time Variability in the Optical Eclipse Depth of WASP-12b}
%\subtitle{Time variability of the optical eclipse depth of WASP-12b}
\author{\mbox{C. von Essen$^{1,2}$}, \mbox{G. Stefansson$^{3,4,5}$}, \mbox{M. Mallonn$^7$}, \mbox{T. Pursimo$^6$}, \mbox{A. A. Djupvik$^6$}, \mbox{S. Mahadevan$^{3,4}$}, \mbox{H. Kjeldsen$^{1,2}$}, \mbox{J. Freudenthal$^8$}, \mbox{S. Dreizler$^8$}} \authorrunning{C. von Essen et al. (2019)}
\titlerunning{New Engineered Diffusers at the NOT}
\offprints{cessen@phys.au.dk}
\institute{$^1$Stellar Astrophysics Centre, Department of Physics and Astronomy, Aarhus University, Ny Munkegade 120, DK-8000 Aarhus C, Denmark\\ 
$^2$Astronomical Observatory, Institute of Theoretical Physics and Astronomy, Vilnius University, Sauletekio av. 3, 10257, Vilnius, Lithuania\\
$^3$Department of Astronomy \& Astrophysics, The Pennsylvania State University, 525 Davey Lab, University Park, PA 16802, USA\\
$^4$Center for Exoplanets \& Habitable Worlds, University Park, PA 16802, USA\\
$^5$NASA Earth and Space Science Fellow\\
$^6$Nordic Optical Telescope, Apartado 474E-38700 Santa Cruz de La Palma, Santa Cruz de Tenerife, Spain \\
$^7$Leibniz-Institut f\"{u}r Astrophysik Potsdam, An der Sternwarte 16, D-14482 Potsdam, Germany\\
$^8$Institut f\"{u}r Astrophysik, Georg-August-Universit\"{a}t G\"{o}ttingen, Friedrich-Hund-Platz\,1, 37077 G\"{o}ttingen, Germany\\
\email{cessen@phys.au.dk}
}

\date{Received ; accepted}

\abstract{We present the characterization of two engineered diffusers mounted on the 2.5 meter Nordic Optical Telescope, located at Roque de Los Muchachos, Spain. To assess the reliability and the efficiency of the diffusers, we carried out several test observations of two photometric standard stars, along with observations of one primary transit observation of \mbox{TrES-3b} in the red (R-band), one of \mbox{CoRoT-1b} in the blue (B-band), and three secondary eclipses of \mbox{WASP-12b} in V-band. The achieved photometric precision is in all cases within the sub-millimagnitude level for exposures between 25 and 180 seconds. Along a detailed analysis of the functionality of the diffusers, we add a new transit depth measurement in the blue (B-band) to the already observed transmission spectrum of \mbox{CoRoT-1b}, disfavouring a Rayleigh slope. We also report variability of the eclipse depth of \mbox{WASP-12b} in the V-band. For the WASP-12b secondary eclipses, we observe a secondary-depth deviation of about 5\,$\sigma$, and a difference of 6\,$\sigma$ and 2.5\,$\sigma$ when compared to the values reported by other authors in similar wavelength range determined from Hubble Space Telescope data. We further speculate about the potential physical processes or causes responsible for this observed variability.}

          
\maketitle

\section{Introduction}
Despite the invaluable efforts in the search for exoplanets carried
out by ground-based observatories such as WASP \citep{Pollaco2006},
HAT-P \citep{Bakos2004}, and KELT \citep{Pepper2007}, the space-based
era of transiting exoplanet searches has completely revolutionized the
field of exoplanets. Lead by the \textit{Kepler} Space Telescope and
its sequential mission K2 \citep{Borucki2010,Howell2014}, the
Convection, Rotation and planetary Transits mission
\citep[CoRoT,][]{Baglin2009} and currently NASA's Transiting Exoplanet
Survey Satellite \citep[TESS,][]{Ricker2015}, these space-based
facilities have shaped our understanding about exoplanets. However,
observatories in space are expensive and generally directly rely on
ground-based follow-up observations to vet and confirm the planetary
nature of exoplanet candidate systems. For example, planets discovered
by TESS closest to the ecliptic will have only $\sim$27 days of
continuous monitoring. In consequence, ground-based facilities will
play a key role, first in their confirmation, and later on, in their
detailed characterization. These follow-up observations will most
likely include a confirmation and characterization of their transits
in different wavelengths \citep[e.g.,][]{Tingley2004,Deeg2009}, a
detailed analysis of their transit timings \citep[see e.g.,][for
  ground-based follow-up of \textit{Kepler} planetary candidates with
  transit timing variations]{vonEssen2018,Freudenthal2018}, and
potentially a characterization of their atmospheres
\citep{Kempton2018}. All of these ground-based follow-up observations
rely on having access to stable and precise photometric data.

Collecting photometry from the ground as precise as space-based data
is a challenging task. Ground-based observations are affected by
scintillation, atmospheric effects that change throughout the
observing nights such as color-dependent absorption of stellar light,
cirrus or clouds passing by, telescope tracking errors, and poor
flat-fielding, among others. These are first-order effects directly
decreasing the quality of the observations, manifesting as a larger
scatter and increased correlated noise in the ground-based light
curves \citep[see e.g.,][]{Smith2006,Carter2009,vonEssen2016}.

To reach precision photometry from the ground to enable robust
characterization of exoplanets, three main alternatives to in-focus
observations have been developed that all rely on spreading out the
light over a number of detector pixels. The basic idea behind the
first technique is to observe with a heavily defocused telescope,
spreading the stellar flux over many pixels over the detector (e.g., a
Charge-Coupled Device, CCD). Spreading the stellar light over many
pixels has the advantage that flat-fielding errors can be better
averaged down compared to focused observations, and further minimizes
the impact of atmospheric seeing. However, the required longer
integration times mean that the sky background level is higher than
for in-focus observations. In addition, by defocusing the telescope
this often creates irregularly illuminated donut-shaped Point Spread
Functions (PSFs) across the detector that change both with seeing and
are susceptible to optical aberrations. Telescope defocusing for
precision photometry was first proposed by \cite{Kjeldsen1992} and
later used e.g., by \cite{Southworth2009} ---and many others--- to
increase the photometric precision of transiting exoplanet light
curves. The second technique relies on using Orthagonal-Transfer CCDs
(OTCCDs) \citep{tonry1997,howell2003}, devices that can directly
shuffle the electrons on the CCD pixels during an exposure to mold a
desired PSF output (e.g., a square). Although this technique has been
used to obtain sub-millimag photometry from the ground \citep[see in
  particular][]{johnson2009} this technique requires OTCCDs which are
expensive and not widely available at different observatories. The
third technique uses Engineered Diffusers (EDs) as a way to scramble
the directions of light that passes through them, spreading the
incoming photons over many pixels on the detector but in a more
homogeneous way than with the defocusing technique
\citep{Stefansson2017,Stefansson2018,Stefansson2018b}. Instead of the
canonically donut-shaped PSF often seen in defocused images, EDs
produce a broad and stabilized PSF with a homogeneous illumination
that is insensitive to seeing effects. At first order, EDs have the
power to increase the exposure times before saturation is reached,
reducing scintillation, and minimizing flat-fielding errors. A full
description of their in-detailed functionality can be found in Section
3.1.4 of \cite{Stefansson2017}. Although the use of EDs for precision
ground-based photometric applications is a new technique, EDs and/or
other structured diffusers have been used for a number of other
applications in astronomy, including direct imaging
\citep{Lafreniere2007} and as light scramblers to provide stable and
homogeneous illumination for fiber-fed spectrographs
\citep[e.g.,][]{halverson2014}. Currently EDs are widely used in
several telescopes for precision photometric applications \citep[see
  Table 1 in][]{Stefansson2018}.

In this work, we present the photometric characterization of two
engineered diffusers mounted at the 2.5 meter Nordic Optical Telescope
(NOT)\footnote{\url{http://www.not.iac.es/instruments/alfosc/diffuser.html}},
henceforth \mbox{ED \#1} and \mbox{ED \#2}. The two EDs have different
diffusing angles to be able to work in crowded and sparse stellar
fields, and with different stellar magnitude ranges for reasonable
exposure times. To characterize them in detail, we have carried out
test observations focused on two photometric standard stars and the
transits of three Hot Jupiters, namely \mbox{TrES-3b}
\citep{ODonovan2007}, \mbox{CoRoT-1b} \citep{Barge2008}, and
\mbox{WASP-12b} \citep{Hebb2009}. Besides the photometric
characterization of the EDs, the collected data allowed us to improve
the transit parameters of \mbox{TrES-3b}, to constrain the
transmission spectrum of \mbox{CoRoT-1b}, and to detect variability of
the eclipse depth of \mbox{WASP-12b}.

The paper is structured as follows. Section~\ref{sec:EDs} presents the
Engineered Diffusers and the basic optical setup at NOT, along with a
description of their overall performance, the presence of ghosts, and
the impact of these in the photometry. Section~\ref{sec:OaDR}
describes the nature of our observations and our data reduction
strategy. Section~\ref{sec:modelling} contains an in-detailed
description of our data modelling, to both primary transits and
secondary eclipses, along with a brief description of our data
detrending. Section~\ref{sec:results} shows our results. We wrap up
this work in Section~\ref{sec:conclusions} with some final remarks.

\section{The Engineered Diffusers at the Nordic Optical Telescope}
\label{sec:EDs}

\subsection{Optical Setup}

For our observations we used the Alhambra Faint Object Spectrograph
and Camera (ALFOSC) and the Filter And Shutter Unit (FASU), where two
extra filter wheels are located in the converging beam in front of
ALFOSC. The detector is a 2k E2V CCD which with the ALFOSC plate scale
of 0.2138\arcsec/pixel gives a FOV of 6.5\arcmin.  The focal length of
the ALFOSC camera is \mbox{131 mm}. Read-out times of ALFOSC for 200
Kpix/sec are 15 seconds in 1$\times$1 binning, and 6.5 seconds in
2$\times$2 binning. The 90 mm diameter diffusers, \mbox{ED \#1} and
\mbox{ED \#2}, were mounted in FASU B, the upper of the two FASU
wheels, with the diffuser coating facing towards the incident beam. We
determined which side has the diffuser coating by looking at the
reflected light coming in at an oblique angle using a torch. Standard
broadband $UBVRI$ and SDSS filters, mounted in a third filter wheel in
the parallel beam inside ALFOSC, or alternatively any filter mounted
in FASU wheel A, can be used in combination with a diffuser in FASU B.

\subsection{The Diffusing Pattern and its Size on Sky}

\begin{figure*}
    \centering
    \includegraphics[width=.3\textwidth]{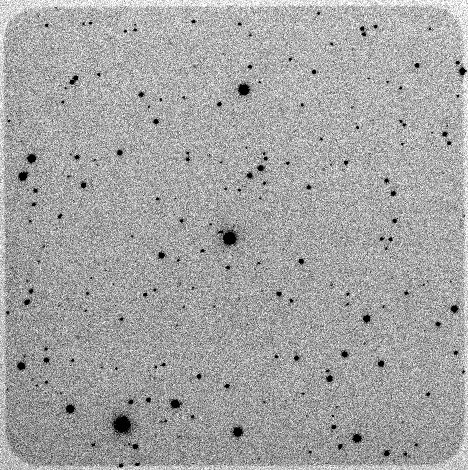}
    \includegraphics[width=.3\textwidth]{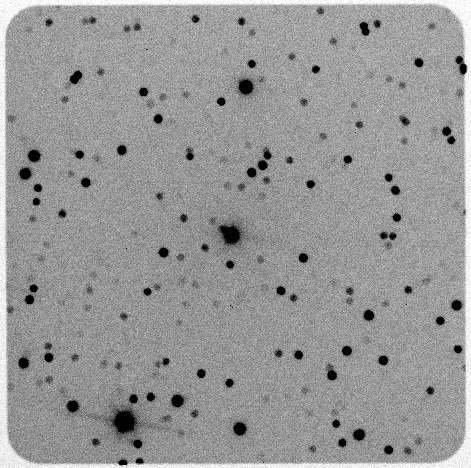}
    \includegraphics[width=.3\textwidth]{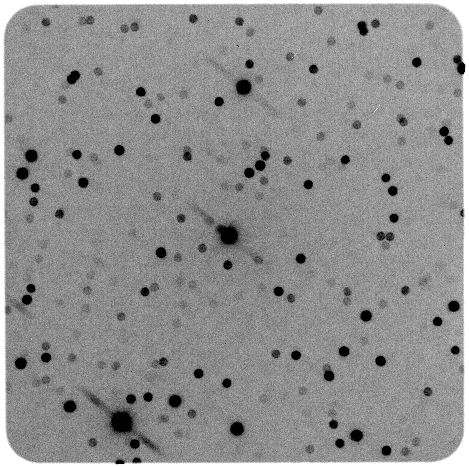}
    \caption{\label{fig:diffusing_pattern} ALFOSC's field of view
      around \mbox{WASP-12}, the brightest star in the center of each
      image. From left to right, without ED, with \mbox{ED \#1} on,
      and with \mbox{ED \#2} on. In the latter, the ghosts described
      in Section~\ref{sec:ghosts} are clearly seen around the
      brightest stars in the field.}
\end{figure*}

The diffusers used in this work were manufactured by RPC Photonics\footnote{\url{https://www.rpcphotonics.com/}} and were funded by the Instrument Center for Danish Astrophysics. Similarly to \cite{Stefansson2017}, the diffusers have a top-hat diffusing pattern and two different diffusing angles of 0.35$^\circ$ and 0.5$^\circ$. To better visualize the effect of the EDs over point sources, Figure~\ref{fig:diffusing_pattern} shows ALFOSC's field of view centered on \mbox{WASP-12} with and without EDs. The diffusing angles differ by a factor of almost two, so the Full-Width at Half-Maximum (FWHM) of the stellar images on the CCD increases about 30\%. The diffusers are coated on one side, which gives a better throughput (about 4\% better by suppressing Fresnel losses). Table~\ref{tab:properties_EDs} summarizes the main properties of the EDs in connection to the NOT, and Table~\ref{tab:specifications} shows their technical specifications as given by \mbox{RPC Photonics}. Even though the diffusing angles differ by a factor of almost two, the FWHM of the stellar images on the CCD increases about 30\%.

\begin{table*}[ht!]
  \caption{\label{tab:properties_EDs} {\it From left to right:} the ED
    number, the diameter (DM) in mm, the diffusing angle, the stellar
    FWHM, for the diffusers placed in FASU A and FASU B, respectively,
    and the distances from the upper filter surface to the focal plane
    (DFP).}  \centering
  \begin{tabular}{l c c c c c c c c c}
    \hline \hline
    ED \#       &    DM          &  Angle           &   FWHM (FASU A)   &   FWHM (FASU B)  &  DFP  \\
                &    (mm)        &  $^{\circ}$      &   (arcsec)        &   (arcsec)       &  (mm)\\
    \hline
    1           &   90           &  0.35            &    4.8            &  5.3             &  100.5 \\
    2           &   90           &  0.50            &    6.4            &  7.5             &  117.0 \\
    \hline
  \end{tabular}
\end{table*}

\begin{table*}[ht!]
\caption{\label{tab:specifications} Technical specifications of the EDs as given by their manufacturers, \mbox{RPC Photonics}.}  
\centering
\begin{tabular}{p{6.5cm} p{10cm}}
\hline \hline
Material & Polymer on replicated on glass substrate\\
Encircled energy & $>$99\% at 2$\times$FWHM (best effort)\\
Max deviation from top-hat irradiance & $<$1\% of spec, $<$3\% requirement in irradiance space, including impact of speckle (best effort) \\
Substrate size & 90 mm +0.05/-0.2 mm diameter\\
Thickness & 3 mm $\pm$0.2 mm thickness (polymer layer adds 0.1 mm to substrate thickness)\\
Anti-reflection coating (unpatterned surface) & Rabs $<$1\%, wavelength 380-1000 nm, AOI 0-5 degrees (best effort) \\
\hline
\end{tabular}
\end{table*}

\subsection{The Stability of the Diffusing Pattern with Airmass and Seeing}

\begin{figure}[ht!]
\centering
\includegraphics[width=.5\textwidth]{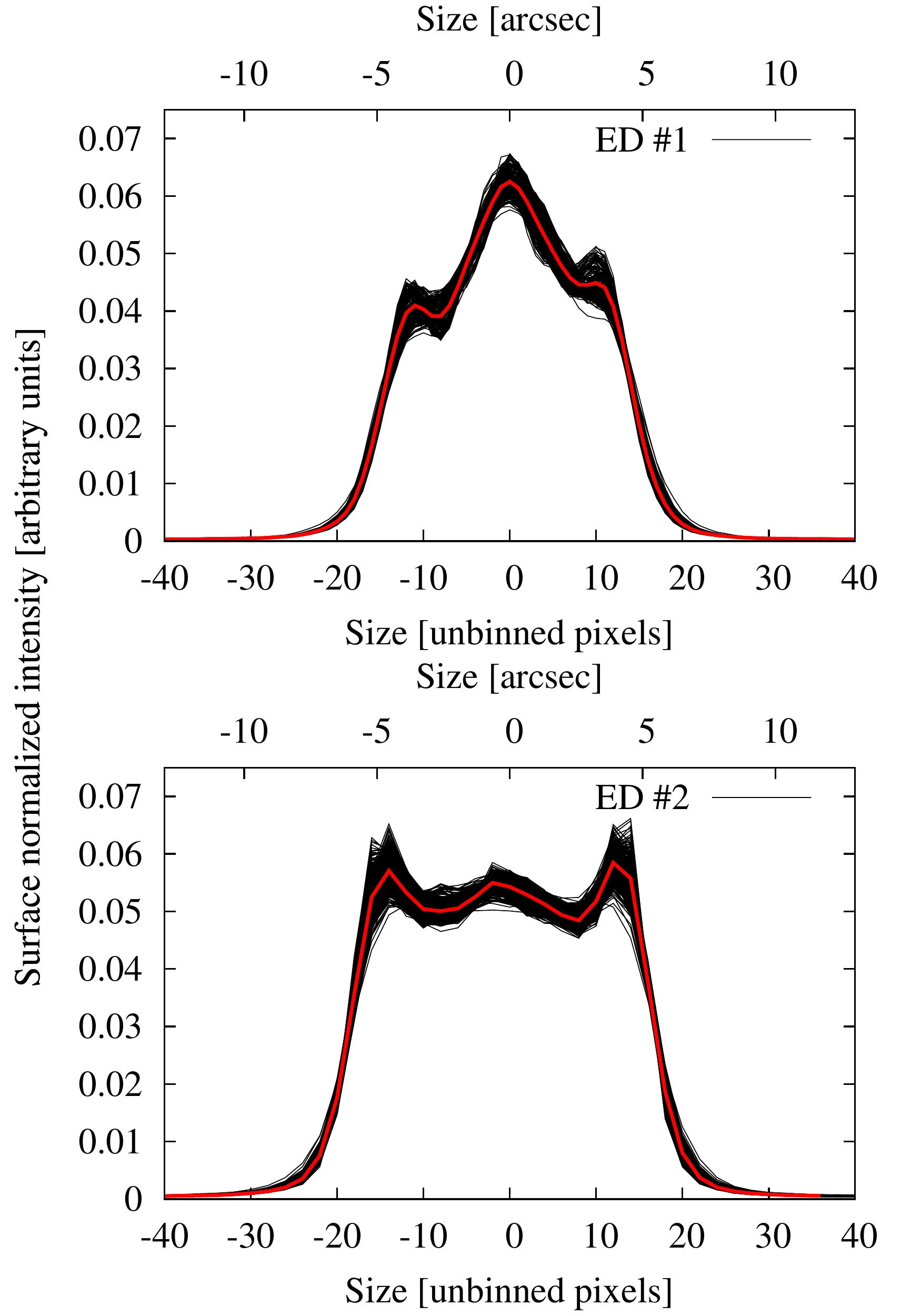}
\caption{\label{fig:diffusing_pattern_2} Diffusing pattern of the EDs
  as a function of unbinned pixels (bottom axis) and arcseconds (top
  axis). The diffused PSF profiles from \mbox{ED \#1} and \mbox{ED
    \#2} are shown in the upper and lower panels, respectively. The
  red line shows the intra-night averaged PSF profile.}
\end{figure}

A key factor in achieving high precision photometry with engineered
diffusers is the stability of the diffusing pattern when the
photometric conditions change during observations as time progresses
\citep[see e.g.,][Figure 3]{Stefansson2017}. To quantify this
stability we calibrated and aligned our ALFOSC diffuser-assisted
observations from January 1$^{st}$ 2019 (184 frames, \mbox{ED \#1})
and January 12$^{th}$ 2019 (473 frames, \mbox{ED \#2}) and computed
normalized East-West cuts of several stars in the ALFOSC field of
view. Figure~\ref{fig:diffusing_pattern_2} shows the resulting
cut-plots for the \mbox{ED \#1} (top) and the \mbox{ED \#2} (bottom).

To quantify the stability of the diffusing pattern, for each pixel
across the East-West cut we computed the difference between the
maximum and minimum values considering the full observing run, and we
divided this difference by its average value. Within the center of the
profile, for the \mbox{ED \#1} the maximum differences are well
contained below 5\% and above 3\%, while for the far wings, where
photometric apertures are usually too large, these values go up to
8\%. For the \mbox{ED \#2}, the center of the profile has a
variability between 3\% and 6\%, while the wings change up to 10\%. As
comparison, we carried out the same exercise to data taken by the NOT
with the telescope that was defocused to achieve a point-spread
function with a FWHM between 2 and 3\arcsec. The observed variability
is well above 20\%.

\begin{figure}[ht!]
\centering
\includegraphics[width=.5\textwidth]{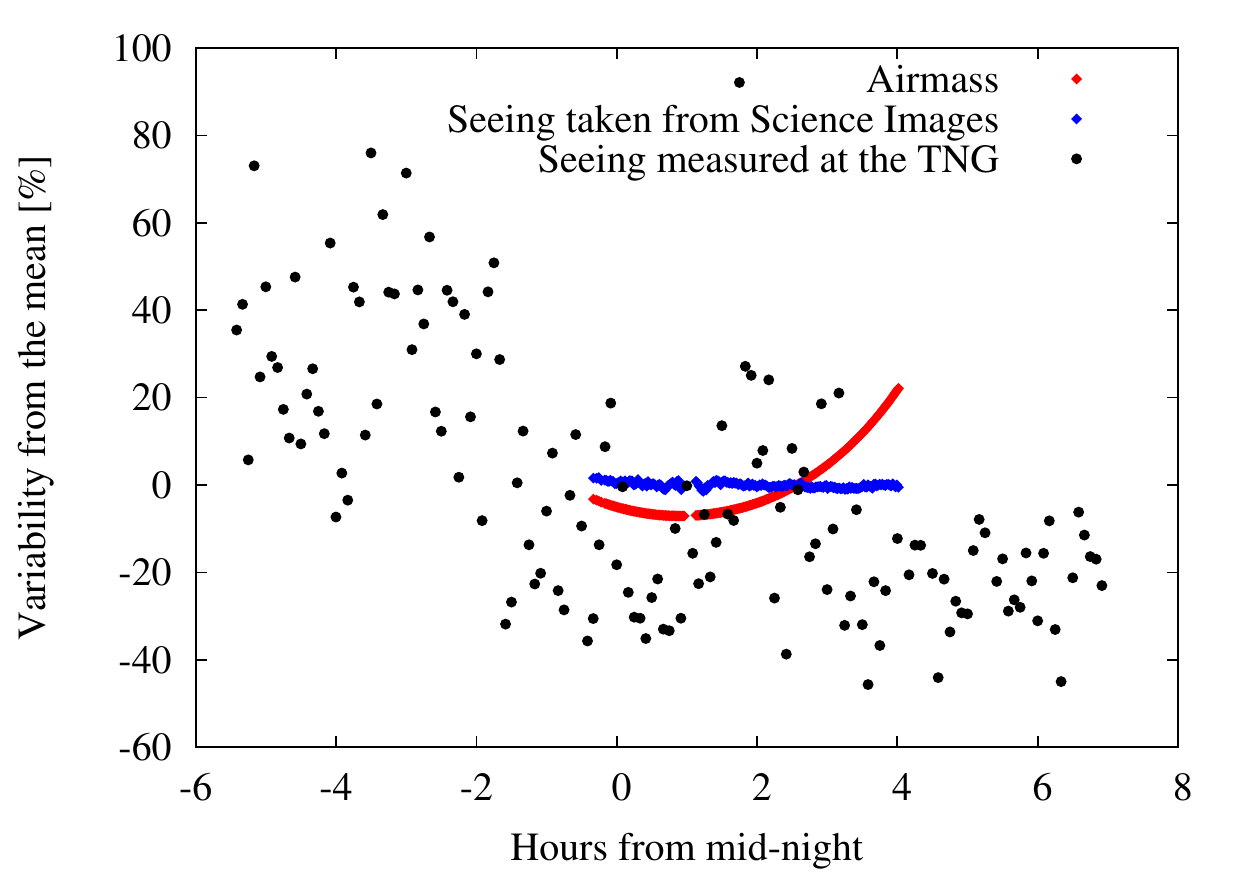}
\caption{\label{fig:AM_S} Variability of airmass (red squares), local
  seeing (black circles) and FWHM of diffuser-assisted science frames
  (blue diamonds) as a function of time within a single night of
  observations. The seeing was calculated by the TNG staff, and the
  diffuser-assisted observations used \mbox{ED \#1} in ALFOSC.}
\end{figure}

To further show the stability of the EDs when seeing and airmass
change along a given observing night, we compared the extent of their
variability to the change in the FWHM of a given set of science
frames. To exemplify this, we have chosen the night where seeing
changed more drastically during the whole night, but specially during
the time range where our observations took
place. Figure~\ref{fig:AM_S} shows the mentioned changes with respect
to their nightly average values for the night of January 1$^{st}$,
2019. To produce the figure, seeing measurements were taken from the
Galileo National Telescope (TNG), airmass was taken from the header of
the images, and the diffused FWHM was computed by averaging the
corresponding FWHMs of several stars within ALFOSC field of view, as
time progressed. The maximum amplitude of variability for the diffused
FWHM corresponds to 3.2\% (blue points in Figure~\ref{fig:AM_S}),
which corresponds to a maximum change in profile of 0.8 unbinned
pixels. On the contrary, the local seeing (black points in
Figure~\ref{fig:AM_S}) changed more than 100\% along the whole night
and 75\% during our observations. A small Pearson's correlation
coefficient between the diffused FWHM and the differential photometry
of -0.2 reflects this stability.

\subsection{Efficiency of the Diffusers and Ghosts}
\label{sec:ghosts}

To determine the throughput of the EDs, on December 2$^{nd}$, 2018,
and on January $20^{th}$, 2019, we observed two standard star fields
found in the \cite{Landolt1992} catalogue, namely \mbox{RU 152} to
characterize the \mbox{ED \#1} in the $U, B, V, R, I$ filters, and
\mbox{SA98 670} for the characterization of the \mbox{ED \#2} in the
$U, B, V, R, I$ and \mbox{SDSS $g \& z$} filters. First, we carried
out a set of $U, B, V, R, I$ images to monitor the zero point of
ALFOSC, followed by the following sequence of observations for each
filter studied: a) filter only, b) filter + Diffuser, c) filter only,
and d) filter + Diffuser, i.e., resulting in 4 images per filter. This
observing strategy allows us to isolate the throughput of the EDs,
effectively correcting for instrument throughput and CCD quantum
efficiency variations, as the measurements were done a few seconds
apart. For this exercise, we typically used the same integration time
as for the zero point monitoring.

\begin{figure}[ht!]
\centering
\includegraphics[width=.5\textwidth]{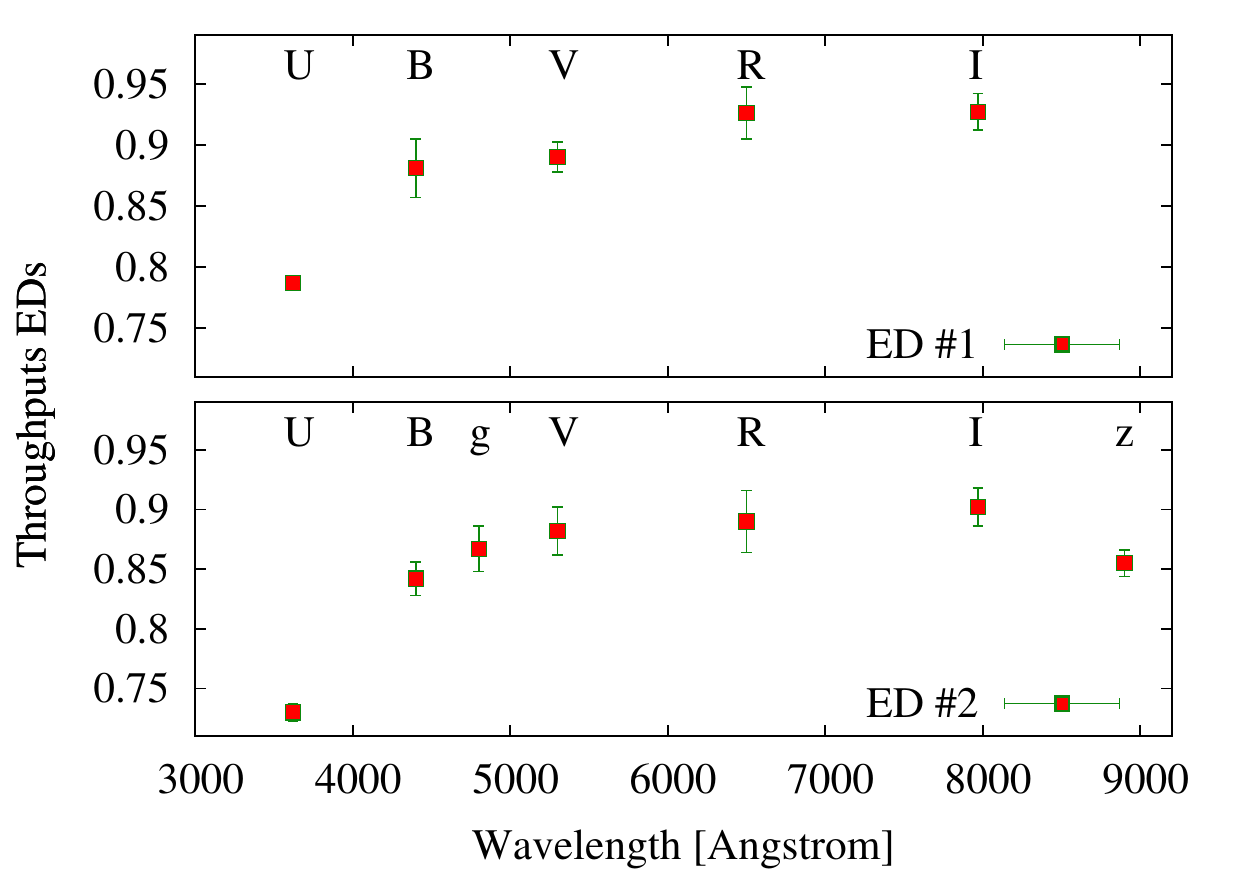}
\caption{\label{fig:throughput_EDs} Throughput of the \mbox{ED \#1} on
  top, and of the \mbox{ED \#2} on bottom. Error bars reflect the
  standard deviation of the measured stellar flux ratios.}
\end{figure}

The bright isolated stars were selected from a zero point frame and
these coordinates were used for the whole sequence. The number of
stars analyzed per filter ranged between 14 and 21 for the \mbox{ED
  \#1}, and 24 and 37 for the \mbox{ED \#2}. In all cases fluxes (in
analog to digital units, ADUs) were measured using IRAF's {\it apphot}
task, with an aperture of 30 pixels (approximately 5.7 arcseconds). In
most cases, the images with and without diffuser have a similar median
count, suggesting that the sky was clear and the night
photometric. Figure~\ref{fig:throughput_EDs} shows the median
throughput for the \mbox{ED \#1} on the top panel, and for \mbox{ED
  \#2} on the bottom panel in the $U, B, V, R, I$ and the $U, B, g, V,
R, I, z$ sets of filters, respectively. The throughput ratios were
computed averaging the flux counts from each set of images for the
respective filters.

From the figures the \mbox{ED \#1} has a throughput of about 90\% in
the B, peaking at 92\% in the R. The U shows a drop placing the
throughput around 80\%. The \mbox{ED \#2} has a similar trend but with
slightly lower values, peaking at 90\% in the red. The drop in the U
band is caused due to the inefficiency of the coating on those
wavelengths (see Table~\ref{tab:specifications}).

\begin{figure}[ht!]
\centering
\includegraphics[width=.5\textwidth]{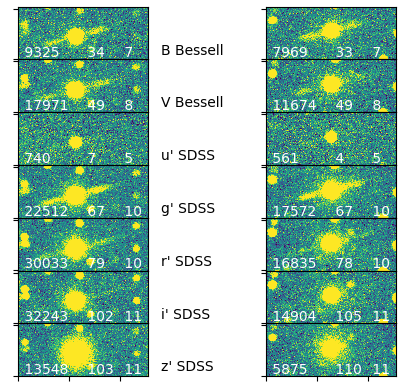}
\caption{\label{fig:ghosts} Stellar intensity profiles where the
  ghosts can be clearly seen. Test images were taken for several
  different photometric filters. {\it Left:} the brightest star is
  close to the edge of the field of view. {\it Right:} the brightest
  star is placed at the center of the field of view. The notation
  "Bessel" corresponds to Bessel filters, and "SDSS" to Sloan
  filters.}
\end{figure}

Furthermore, from our test images taken with the \mbox{ED \#2}, we
noted some structure around the brightest stars, specially when the
count peaks were above $\sim$60\% the total dynamic range of ALFOSC's
CCD. We call these features ghosts. Figure~\ref{fig:ghosts} shows the
ghosts near bright stars that were captured using different
filters. The two sided ghost appear to be stronger in bluer
wavelengths than in redder ones. When the star is at the centre of the
field of view, the left hand side ghost is stronger than the
right. However, this is reversed when the star is close to the edge of
the field of view. To produce the figure, the data were overscan
subtracted, but not flat field corrected. To estimate the relative
intensity between the stellar flux and the ghosts, we measured their
respective count levels using IRAF task {\it imexamine}. While stellar
intensities are given by the counts at the peak of the PSFs, the
intensity of the ghosts are estimated averaging a 5$\times$5 pixel
square near the maximum of the structure, determined from visual
inspection. The ratio between the stellar intensity peak and the
averaged ghost counts is approximately 196 for the $U$, 296 for the
$B$, 550 for the $V$, 1290 for the $R$, 3500 for the $I$, and 290 for
the $g$, revealing the increase of its contribution towards bluer
wavelengths. The z seems to show no ghosts. The ratios are in all
cases so large, that we do not expect --and do not observe-- that
these features will have any impact in the quality of the photometry.

\section{Observations and Data Reduction}
\label{sec:OaDR}

\subsection{The Collected Data}

%http://telescope.livjm.ac.uk/Reports/

In Table~\ref{tab:obscond} we summarize the main characteristics of all our science observations. 

On the night of August, 16$^{th}$, 2018, we observed one primary
transit of \mbox{TrES-3b} with \mbox{ED \#1}. The host star has a
magnitude of \mbox{R = 11.98} \citep{ODonovan2007}, making it a
relatively bright source for the collecting area of the NOT. The
transit is fully covered, with about 15 minutes of out-of-transit data
before and after the transit. The night was stable and clear. The data
were taken in 1x1 binning using a Bessel $R$ filter, centered at
\mbox{650 $\pm$ 130 nm}. The overall, averaged cadence, accounting for
readout time, was 36 seconds, while the exposure time was set to 25
seconds. The night was photometric and with good seeing, slowly and
steadily increasing between 0.5 and 1.5 arcseconds along the night.

On the night of December, 11${th}$, 2018, we observed one primary
transit of \mbox{CoRoT-1b} using the \mbox{ED \#1}. The transit
coverage was complete, with 45 minutes and 1 hour of out-of-transit
data before and after the transit, respectively. The magnitude of the
host star is \mbox{V = 13.6}, and its spectral type is G0V
\citep{Guenther2012}. We observed the host star and several comparison
stars during primary transit in 1x1 binning using a Bessel $B$ filter
\mbox{(440 $\pm$ 100 nm)}. To the best of our knowledge, these data
correspond to the bluest transit observations of \mbox{CoRoT-1b}
carried out to date. Due to the magnitude of the host star, the
overall cadence was 180 seconds. The seeing varied in the night
between 0.5 and 2\arcsec, but most of the time well below
1\arcsec. The night was overall photometric.

Finally, on the nights of January 1$^{st}$, 12$^{th}$, and 24$^{th}$,
2019, we observed three secondary eclipses of \mbox{WASP-12b}, the
first and last one using the \mbox{ED \#1} and the one in between
using the \mbox{ED \#2}. In all cases the secondary eclipses are fully
covered, with approximately 20 minutes to 1 hour of out-of-eclipse
data before and after the events. Following the given dates, the
overall cadences were between 43, 37 and 87 seconds. All eclipses were
observed with the Bessel $V$ filter, with the main goal to
characterize the geometric albedo of the exoplanet (see
Section~\ref{sec:WASP12b} for further details).

\begin{table*}[ht!]
  \caption{\label{tab:obscond} Most relevant parameters describing our
    observations. {\it From left to right}: The date corresponding to
    the beginning of the local night in years, months, and days
    (yyyy.mm.dd), the name of the target, the filter in which the
    observations were performed, the ED used for the observations, the
    number of data points per light curve, N, the average cadence in
    seconds, CAD (accounting for readout time), the exposure time, ET,
    in seconds, the total observing time in hours, T$_{tot}$, and the
    airmass range, $\chi_{min,max}$, showing minimum and maximum
    values.}  \centering
  \begin{tabular}{l c c c c c c c c}
    \hline \hline
    Date       &    Name           &  Filter      &    ED \#  &  N    &   CAD      &   ET         &   T$_{tot}$ &   $\chi_{min,max}$  \\
    yyyy.mm.dd &                   &              &           &       &  (sec)     &   (sec)      &    (hours)  &                     \\
    \hline
    2018.08.16 &   TrES-3b         &  Bessel $R$  &    1      & 167   &     36     &   25         &  1.67       &  1.00 - 1.08 \\
    2018.12.11 &   CoRoT-1b        &  Bessel $B$  &    1      &  82   &    180     &  162         &  4.10       &  1.00 - 1.51 \\
    2019.01.01 &   WASP-12b        &  Bessel $V$  &    1      & 184   &     84     &   50         &  4.30       &  1.00 - 1.31 \\
    2019.01.12 &   WASP-12b        &  Bessel $V$  &    2      & 473   &     37     &   25         &  4.92       &  1.00 - 1.34 \\
    2019.01.24 &   WASP-12b        &  Bessel $V$  &    1      & 202   &     87     &   60         &  4.91       &  1.00 - 1.49 \\            
    \hline
  \end{tabular}
\end{table*}

\subsection{Data Reduction}

The data reduction is carried out by means of our {\it Differential
  Photometry Pipelines for Optimum Lightcurves}, DIP$^2$OL. The
procedure is fully described in \cite{vonEssen2018}. In brief, the
photometric reduction pipeline is based in IRAF's command language,
and carries out normal calibration sequences such as bias and dark
subtraction and flatfielding, using the task {\it ccdproc}. The
science frames are then scrutinized for cosmic rays using IRAF's task
{\it cosmicrays}. Afterwards, the calibrated images are aligned using
{\it xregister}, and counts within several combinations of different
apertures and sky rings are computed. The apertures are distributed
between 0.5 and 5 times the average stellar PSF FWHM of the night
computed directly from the science frames, while the sky rings between
1 and 3. As the PSFs are diffused, it is worth to mention that the
seeing we compute is absolutely not representative of the local seeing
of the site. For the data detrending (see Section~\ref{sec:modelling})
DIP$^2$OL outputs the airmass, the seeing, the (x,y) pixel coordinates
of the centroid positions, the background counts determined within the
sky rings, and the the integrated master flatfield and master dark
counts for each aperture, when available. The last three quantities
are computed per measured star. Our data reduction finishes converting
the time axis given in Julian dates to Barycentric Julian dates. To do
so, we make use of the web
tool\footnote[1]{http://astroutils.astronomy.ohio-state.edu/time/utc2bjd.html}
presented in \cite{Eastman2010}. For this, we use as input values the
geographic coordinates of the NOT, along with its altitude above sea
level, and the celestial coordinates of the stars.

\section{Data Modelling}
\label{sec:modelling}

\subsection{Error Determination for the Fitting Parameters}

To determine reliable errors for the parameters fitted in this work,
we sample from the posterior-probability distribution using
Markov-chain Monte Carlo (MCMC). Our MCMC routines are python-based
and make use of
\texttt{PyAstronomy}\footnote{\url{http://www.hs.uni-hamburg.de/DE/Ins/Per/Czesla/}\\ \url{PyA/PyA/index.html}}
\citep[based on PyMC,][]{Patil2010,Jones2001}. To compute our best-fit
parameters we iterated \mbox{100\ 000} times per transit and eclipse,
and discarded a conservative first 20\% of the samples as burn-in. To
investigate the convergence of the chains we first visually inspected
them, and then we divided them in three equal parts, computing the
best-fit parameters and their errors in each sub-chain, and verifying
consistency among results. In this work, error bars are given at
1-$\sigma$ level (68.3\% credibility intervals).

\subsection{Primary Transit}

As the primary transit model we used the one given by
\cite{MandelAgol2002}. From the transit light curve we can infer the
orbital period, P, the mid-transit time, $\mathrm{T_0}$, the
planet-to-star radii ratio, $\mathrm{R_p/R_s}$, the distance between
planet and star centers in units of stellar radii, $\mathrm{a/R_s}$,
and the orbital inclination, i, in degrees. We modelled the stellar
limb-darkening with a quadratic law, with corresponding limb-darkening
coefficients, $\mathrm{u_1}$ and $\mathrm{u_2}$. The latter were taken
from \cite{Claret2011} for the used filters, and closely matching the
effective temperature, the metallicity and the surface gravity of each
one of our observed stars. For the fitting parameters we always
considered uniform priors using as starting values the ones obtained
from the literature. The width of the priors was set to
$\pm$30-$\sigma'$, being $\sigma'$ the errors reported by the
respective authors.

\subsection{Secondary Eclipse}

As secondary eclipse model we used a scaled version of
\cite{MandelAgol2002} transit model. In this case, both linear and
quadratic limb-darkening coefficients were set to zero. If the transit
model is represented by TM(t), then the secondary eclipse model,
SEM(t) has the following expression:

\begin{equation}
\mathrm{SEM(t)} = (\mathrm{TM(t)} - 1.) \times \mathrm{sf} + 1
\end{equation}

\noindent In the equation, sf denotes the scaling factor. This
approach conserves perfectly the transit duration and shape. The
eclipse depth is then computed as $(\mathrm{R_P/R_s})^2 \times
\mathrm{sf}$, and its error computed from error propagation.

\subsection{Data Detrending}

Our detrending strategy is fully described in \cite{vonEssen2018}. In
brief, for the detrending model we consider a linear combination of
seeing, airmass, centroid positions of the stars involved in the
differential photometry, integrated counts over flats and darks on the
selected aperture, when available, and integrated sky counts for the
selected sky ring. To choose the combination of parameters required to
properly detrend the data without over-fitting them, we take into
consideration the joint minimization of four statistical indicators:
the reduced-$\chi^2$ statistic, the Bayesian Information Criterion,
the standard deviation of the residual light curves enlarged by the
number of fitting parameters, and the Cash statistic. During our
fitting procedure, at each iteration the detrending coefficients are
computed from simple inversion techniques, considering the randomly
chosen primary transit or secondary eclipse parameters as fixed. Then,
the combined model of detrending times primary transit/secondary
eclipse models undergo the $\chi^2$ minimization associated to our
MCMC strategy. In consequence, the uncertainties of the detrending
coefficients propagate into the uncertainties of the primary transit
and secondary eclipse fitting coefficients.

In addition to the detrending strategy produced in
\cite{vonEssen2018}, here we add a further checkup to verify that the
detrending does not have any impact of significance in the derived
fitting parameters. In the case of the secondary eclipses, where the
depth is intrinsically low and thus data might suffer most from
over-fitting, once we determined our best-fit secondary eclipse depth
we verified its proper convergence by means of a $\chi^2$ map. For the
map, we created a grid of secondary eclipse depths within a reasonable
range. We assumed that each one of these values were the ones
reproducing our data best, and we computed the coefficients of the
linear combination of the detrending model. With these assigned, we
computed $\chi^2$ and plotted its distribution as a function of the
eclipse depth. If the detrending process was properly produced and the
MCMC chains converged to their absolute minima, the $\chi^2$ map
should have a unique minimum around the best-fit value. The resulting
$\chi^2$ maps are given in Section~\ref{sec:results}.

\subsection{Noise Treatment}

To test the extent to which our data are affected by correlated noise
\citep{Pont2006}, we followed the approach described in
\cite{Carter2009} and \cite{vonEssen2018}. After carrying out the MCMC
fit, we computed the residual light curves by dividing the photometric
data by the best-fit model, and from the residuals we computed the
$\beta$ factor. Briefly, the residuals are divided into M bins of N
averaged data points. This average accounts for changes in exposure
time that might be needed to compensate for changes in airmass or
transparency during the observing runs. If the data show no correlated
noise, then the noise contained in the residual light curves should
follow the expectation of independent random numbers:

\begin{equation}
  \hat{\sigma}_N = \sigma N^{-1/2}[M/(M-1)]^{1/2}\ .
\end{equation}

\noindent where $\sigma$ is the standard deviation of the unbinned
residual light curve, while $\sigma_N$ corresponds to the standard
deviation of the data binned with N averaged data points per bin:

\begin{equation}
  \sigma_N = \sqrt{\frac{1}{M}\sum_{i = 1}^{M}(<\hat{\mu}_i> - \hat{\mu}_i)^2}\ .
\end{equation}

\noindent Here, $\hat{\mu_i}$ corresponds to the mean value of the
residuals per bin, i, and $<\hat{\mu_i}>$ is the mean value of the
means. If some amount of correlated noise is present in the data, then
$\sigma_N$ and $\hat{\sigma}_N$ should differ by a factor, called
$\beta_N$. To compute the $\beta$ value, we have averaged all the
$\beta_N$'s obtained from bins that are as large as 0.8, 0.9, 1, 1.1
and 1.2 times the duration of ingress. When $\beta$ was found to be
larger than 1, we have enlarged the individual photometric error bars
by this factor, and carried out the MCMC fitting process again.

\section{Results}
\label{sec:results}

\subsection{TrES-3b}

\begin{figure}[ht!]
    \centering
    \includegraphics[width=.5\textwidth]{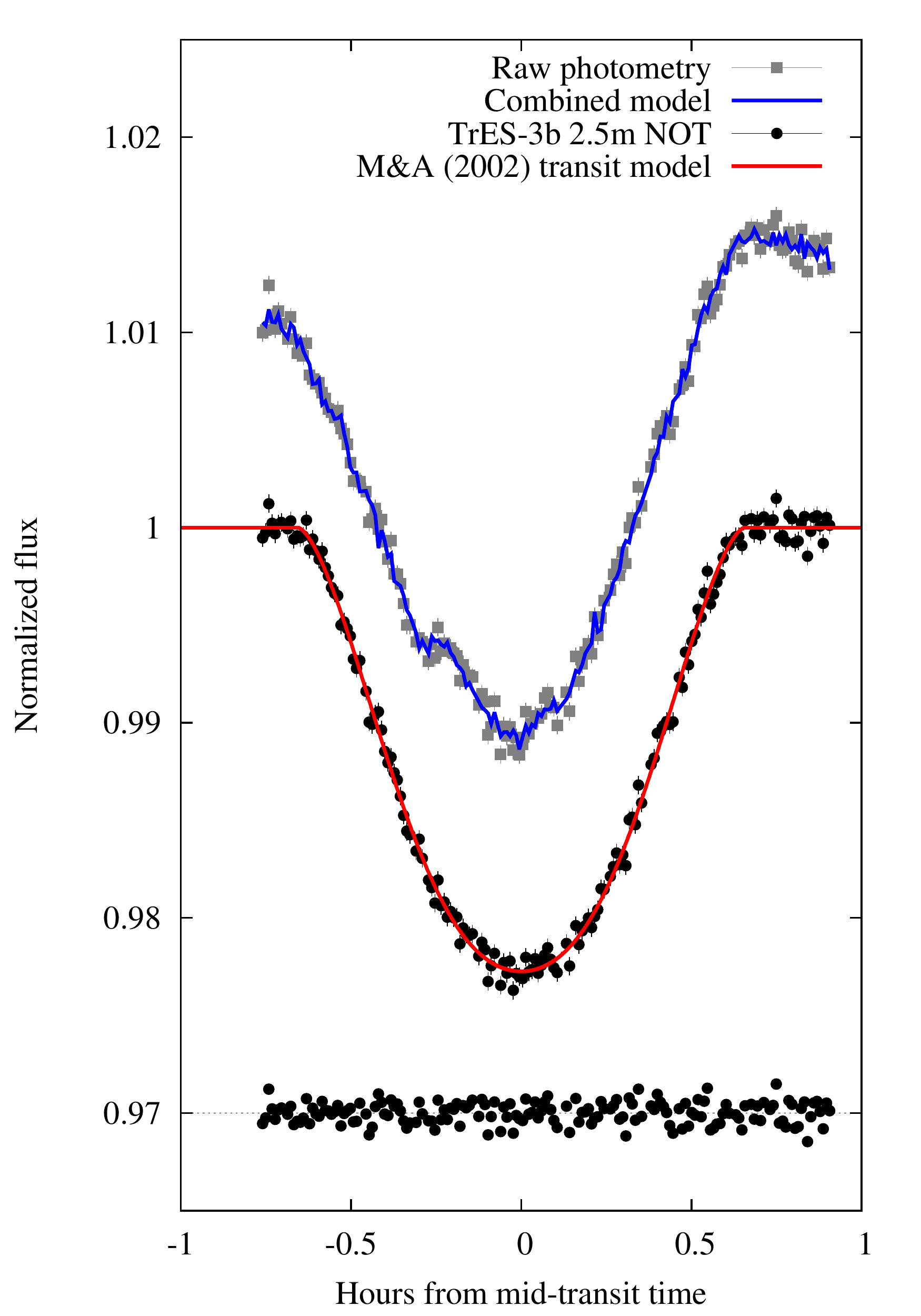}
    \caption{\label{fig:TrES-3} Primary transit photometry for
      \mbox{TrES-3b}, as a function of the hours from the best-fit
      mid-transit time. From top to bottom, raw photometry in gray
      squares, along with the combined best-fit model (transit times
      detrending) in blue continuous line. Black circles and error
      bars correspond to the detrended photometry, and the red
      continuous line to the best-fit transit model. Residuals are
      shown below. All quantities are artificially shifted to allow
      for visual inspection.}
\end{figure}

To determine our best-fit parameters from the data collected during
the primary transit of \mbox{TrES-3b}, as starting values for the
transit parameters we used the ones found in \cite{ODonovan2007},
their Table 3. Since we have only one transit light curve, the orbital
period was considered as fixed using the value reported in the
literature. We then fitted the mid-transit time, the distance between
planet and star centers in units of stellar radii, the inclination,
and the planet-to-star radii ratio. Our derived results, along with
1-$\sigma$ error bars, are listed in Table~\ref{tab:TrES-3}. Our
values are fully consistent with the ones reported by
\cite{ODonovan2007}, with smaller
uncertainties. Figure~\ref{fig:TrES-3} shows the raw data in grey
squares, the combined model in blue continuous lines, the detrended
data in black circles, and the best-fit transit model in red
continuous line. The standard deviation of the residual light curve
for an exposure time of 25 seconds is 0.5 parts-per-thousand (ppt),
and the corresponding $\beta$ value is very close to 1. It is worth to
mention that this target was also observed by \cite{Stefansson2017}
using their own ED with a similar diffusing angle (0.34 versus 0.35
degrees). For their test observations they used the 3.5 meter
Astrophysical Research Consortium Telescope, located at Apache Point
Observatory, United States of America. Even though their collecting
area is significantly larger than the one of the NOT, their achieved
precision is 0.75 ppt at a 32.5 second cadence. To our knowledge, the
only other ground-based transit light curve of \mbox{TrES-3b} more
precise than the one reported here can be found in
\cite{Parviainen2016}. The authors observed \mbox{TrES-3b} with the 10
meter Gran Telescopio Canarias and the OSIRIS low-resolution
spectrograph, achieving in their white light curve a 0.5 ppt scatter
for 12 seconds exposure.

\begin{table}[ht!]
    \caption{\label{tab:TrES-3} Transit parameters (TPs) for
      \mbox{TrES-3b}. From top to bottom the orbital period in days,
      the mid-transit time in Barycentric Julian Dates, the semi-major
      axis in stellar radius, the inclination in degrees, the
      planet-to-star radii ratio, and the linear and quadratic limb
      darkening coefficients. The mid-transit time is given in
      BJD$_{TDB}$ minus 2454000.}
    \begin{tabular}{l c c}
    \hline\hline
    TPs                  &  \cite{ODonovan2007}  &   This work      \\
    \hline
    P (days)             & 1.30619 $\pm$ 0.00001  & adopted         \\
    T$_0$ (BJD$_\mathrm{TDB}$) & 185.9101 $\pm$ 0.0003 & 4347.4216 $\pm$ 0.0002 \\
    a/R$_s$              &   6.06 $\pm$ 0.10      &  6.07 $\pm$ 0.06 \\
    i ($^{\circ}$)        &   82.15 $\pm$ 0.21    & 81.93 $\pm$ 0.13 \\ 
    R$_p$/R$_s$          &   0.1660 $\pm$ 0.0024  & 0.1614 $\pm$ 0.0036 \\
    u$_1$, u$_2$         &            -            & 0.3612, 0.2814 \\
    \hline
    \end{tabular}
\end{table}

\subsection{CoRoT-1b}

For \mbox{CoRoT-1b}, as initial values for our MCMC fit we used the
parameters found in \cite{Bean2009}, their Table 1 and 3, which in
turn were derived from a global fit of 36 transit light curves taken
with the CoRoT space based telescope, 20 of them taken in long cadence
(512 second exposures corresponding to a 0.681 ppt photometric
precision) and the remaining ones taken in CoRoT's "short runs"
\citep[32 seconds with a photometric precision of 2.724 ppt, see][for
  the definition of long cadence and short run]{Baglin2006}.

\begin{figure}[ht!]
    \centering
    \includegraphics[width=.5\textwidth]{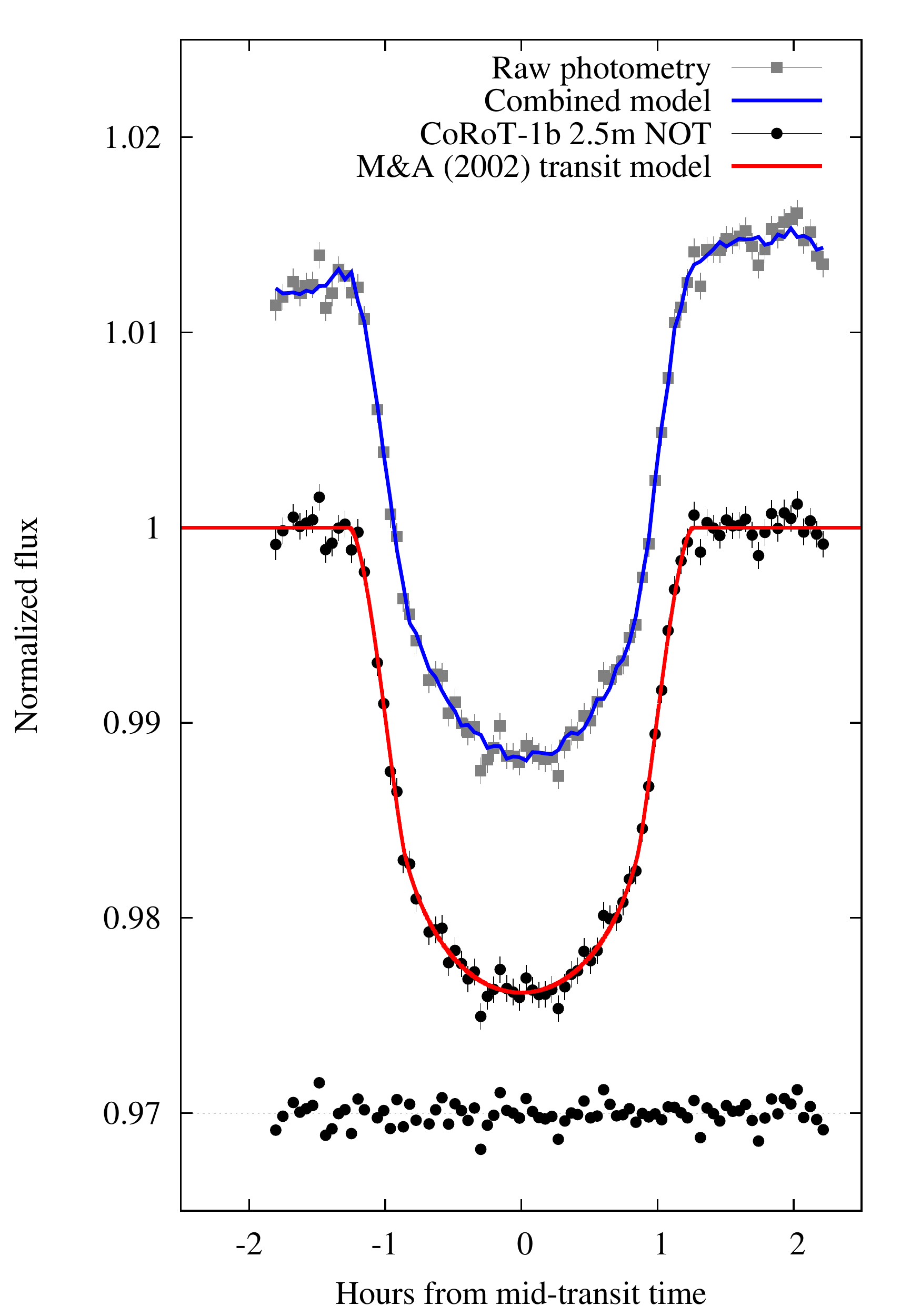}
    \caption{\label{fig:CoRoT-1} Primary transit photometry of
      \mbox{CoRoT-1b}, as a function of the hours from the best-fit
      mid-transit time. The figure description is the same as the one
      given in Figure~\ref{fig:TrES-3}.}
\end{figure}

In Table~\ref{tab:CoRoT-1} we list our best-fit transit parameters
along with 1-$\sigma$ errors, compared to the ones reported by the
authors. All values are consistent within uncertainties to the ones
reported by \cite{Bean2009}. Figure~\ref{fig:CoRoT-1} shows the
detrended transit photometry in black circles, along with our best-fit
transit model in red continuous line. The standard deviation of the
residual light curve, for an exposure time of 162 seconds, is 0.65
ppt. The corresponding $\beta$ value is close to 1. Considering that
our exposure time is three times smaller than CoRoT's long cadence,
our derived photometric precision is significantly superior than
CoRoT's.

\begin{table}[ht!]
    \centering
    \caption{\label{tab:CoRoT-1} Transit parameters (TPs) for
      CoRoT-1b, following the same description as
      Table~\ref{tab:TrES-3}.}
    \begin{tabular}{l c c}
    \hline\hline
    TPs                  &  \cite{Bean2009}  &   This work      \\
    \hline
    P (days)             & 1.5089656 $\pm$ 0.0000060        & adopted \\
    T$_0$* (BJD$_\mathrm{TDB}$) & 159.452879 $\pm$ 0.000068 & 4464.54058 $\pm$ 0.00018 \\
    a/R$_s$              &  4.751 $\pm$ 0.045               &  4.751 $\pm$ 0.036 \\
    i ($^{\circ}$)       &  83.88 $\pm$ 0.29                &  83.96 $\pm$ 0.18 \\ 
    R$_p$/R$_s$          &  0.1433 $\pm$ 0.0010             &  0.1419 $\pm$ 0.0019 \\
    u$_1$, u$_2$         &       -                          &   0.5700, 0.2165 \\
    \hline
    \end{tabular}
\end{table}

\begin{figure}[ht!]
    \centering
    \includegraphics[width=.5\textwidth]{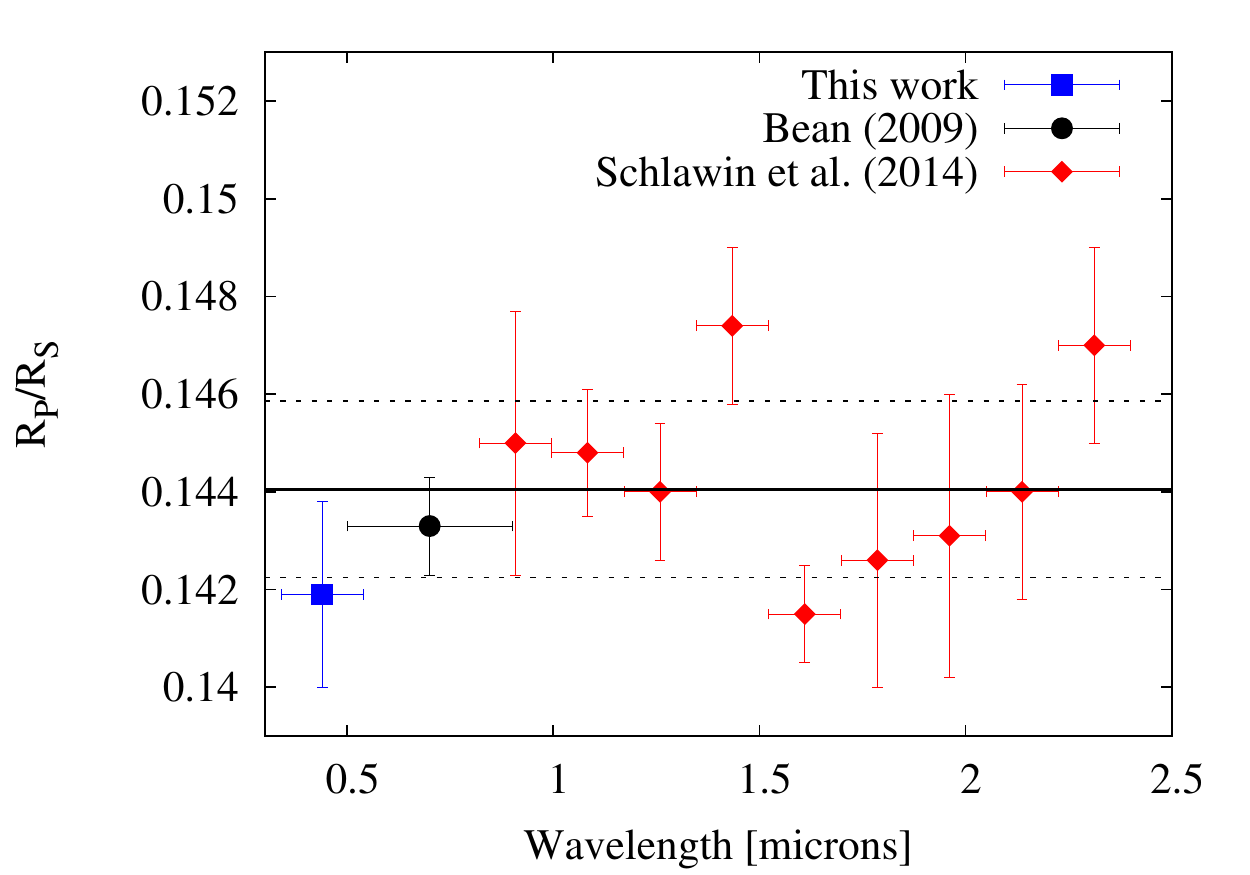}
    \caption{\label{fig:CoRoT1_TS} Transmission spectrum of
      \mbox{CoRoT-1b}, with planet-to-star radii ratio obtained from
      this work (blue square), \cite{Bean2009} (black circle), and
      \cite{Schlawin2014} (red diamonds). Horizontal continuous and
      dashed lines indicate the mean and $\pm$ the standard deviation
      of the data points, respectively.}
\end{figure}

As previously mentioned, our data were taken with the \mbox{ED \#1} in
simultaneous to the Bessel $B$ filter. In consequence, our derived
planet-to-star radii ratio is, to date, the bluest point characterized
in \mbox{CoRoT-1b's} transmission spectrum. Figure~\ref{fig:CoRoT1_TS}
shows its complete transmission spectrum. A non-chromatic planet-star
radius ratio results in a $\chi^2$ of 15.79 with 11 degrees of freedom
(equivalently, a reduced $\chi^2$ of 1.43) and a probability $P$ of
0.1. Thus, the current set of observations do not show significant
variations in R$_p$/R$_s$. \mbox{CoRot-1b} has an atmospheric pressure
scale height of $\sim$\,700~km. Models of hot Jupiter transmission
spectra predict variations of $\mathrm{R_P/R_S}$ in low spectral
resolution within approximately $\pm$~2 scale heights
\citep{Fortney2010,Schlawin2014}. The mean uncertainty of the existent
measurement for $\mathrm{R_P/R_S}$, and also the uncertainty of the
new measurement of this work, roughly correspond to this value of two
scale heights. In consequence, the data are not precise enough for a
detailed transmission model fitting. Nonetheless, the new $B$ band
data point is lower than the mean by $\sim$\,2.5~scale heights. The
optical transmission spectra of many other Hot Jupiters show an
increase of the $\mathrm{R_P/R_S}$ values towards short wavelengths by
about a few scale heights due to scattering
\citep{Sing2016,Mallonn2017}. Thus, the observations reported here
disfavor a typical scattering signature in the transmission spectrum
of \mbox{CoRoT-1b}. If confirmed by follow-up observations, it might
be indicative for a grey absorption by clouds composed of rather large
condensate particles \citep{Wakeford2015}.

\subsection{WASP-12b}
\label{sec:WASP12b}

The three eclipses of WASP-12b were fitted independently from each
other, using as fitting strategy the one described in previous
sections. For the scaling factor of the primary transit we allowed
both positive (physical) and negative (nonphysical) values. To fit the
scaling factor we considered uniform priors corresponding to a
reasonable secondary eclipse depth of $\pm$3 ppt. For the transit
parameters we adopted the values recently reported by
\cite{Maciejewski2018}, particularly for an orbital period that allows
for the observed orbital decay. The mid-eclipse times for our observed
secondary eclipses, $\mathrm{T_{o,SE}}$, were computed from a
quadratic ephemeris:

\begin{equation}
    \mathrm{T_{o,SE}} = \mathrm{T_o} + \mathrm{P} \times \mathrm{E} + \frac{1}{2}\mathrm{\frac{dP}{dE}} \times \mathrm{E^2}\,.
\end{equation}

\begin{figure*}[ht!]
    \centering
    \includegraphics[width=.85\textwidth]{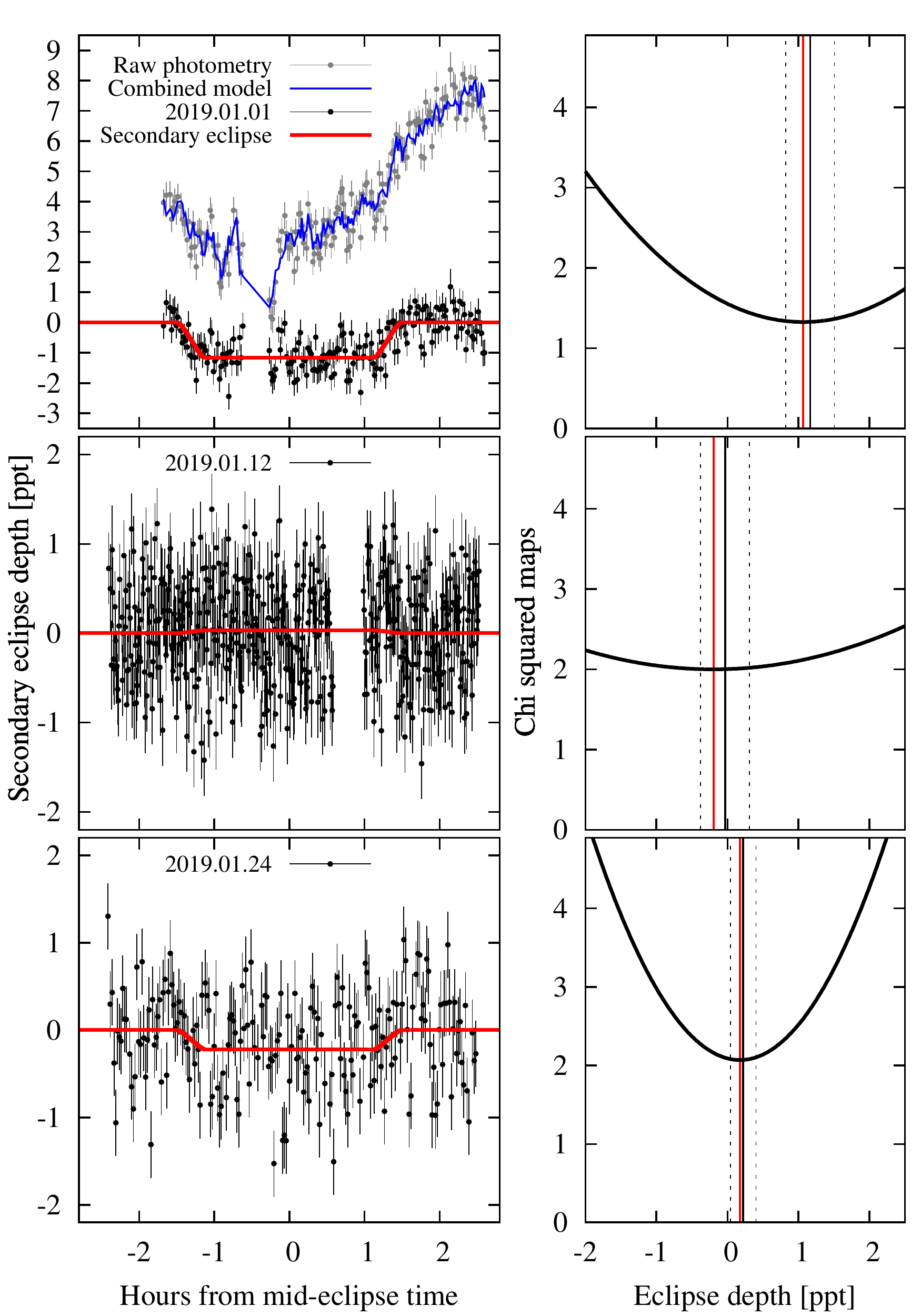}
    \caption{\label{fig:WASP12_sec_eclipse_data} {\it Left:} secondary
      eclipse data of \mbox{WASP-12b} obtained with the NOT and the
      EDs on. Black circles and error bars correspond to the
      photometric data. The red continuous line shows the best-fit
      secondary eclipse, in ppt. For the most prominent eclipse we
      show raw photometry in gray squares, along with the combined
      best-fit model (transit times detrending) in blue continuous
      line. {\it Right:} $\chi^2$ maps for different scaling
      factors. The thick black line corresponds to the $\chi^2$ values
      obtained for different eclipse depths. The red vertical line
      indicates the minimization of $\chi^2$, and the black continuous
      vertical line, along with the two dashed ones, indicate the
      best-fit MCMC results and a $\pm$2-sigma contour.}
\end{figure*}

\noindent E corresponds to the epoch of the observation with respect
to the mid-transit time of reference, $\mathrm{T_o}$. The values of
$\mathrm{T_o}$, P and $\mathrm{\frac{dP}{dE}}$ can be found in
Table~\ref{tab:WASP-12}, along with other relevant transit parameters
fully adopted in this work. In Table~\ref{tab:WASP-12_SEDs} we report
the best-fit secondary eclipse depths for the three dates, along with
other information that reflects the quality of our data and fit. The
left panels of Figure~\ref{fig:WASP12_sec_eclipse_data} shows our
resulting secondary eclipse light curves for the mentioned nights, and
the right panels show our $\chi^2$ maps for each one of them, compared
to the MCMC best-fit values and errors at a 2-$\sigma$ level. In all
cases, the detrending parameters that were always favoured were the
x,y centroid positions of the stars and the airmass. The first set of
parameters does not come as a surprise, as the telescope was passing
very close to the zenith, and its there where the stability of the
tracking is not optimal. Because of this, a $\sim$20 minute gap is
clearly seen during the first and second nights. This data gap is
mostly produced by our 5-$\sigma$ clipping algorithm that takes place
during the construction of the raw light curves, as these data points
naturally have large photometric uncertainties. Originally, some
sparse data points were left within the gap, but we noticed that
including them in the fitting procedure made the number of detrending
parameters reach their maximum to account for this noise. In
consequence, to minimize the number of detrending components we
discarded the few points left inside the gap, and carried out our
usual fitting procedure with the minimum possible detrending
components. Finally, extinction of light crossing our atmosphere is an
effect that is inherent from the diffusers, and thus is found in our
detrending components as well.

% epoch 3643.5 01.01.2019, 2458485.56555
% epoch 3653.5 01.12.2019, 2458496.47974
% epoch 3665.5 24.01.2019, 2458508.48534

\mbox{WASP-12} is a triple star system with a separation of only
1\arcsec between the primary component and its low-mass companions
\citep{Bechter2014}. In the $V$ band, the flux contribution of both
companions amounts to about 0.9\% \citep{Crossfield2012} and is fully
included in our photometric apertures. The measured eclipse depth of
Table~\ref{tab:WASP-12_SEDs} are diluted by this amount and need to be
corrected by 0.9\% upwards. However, we report the uncorrected,
diluted values because the correction is one order of magnitude
smaller than the given uncertainties.

The depth of the first eclipse measurement deviates by about
5\,$\sigma$ from the depth of the second and third event. The latter
are in agreement at the 2\,$\sigma$ level. Observations of the
secondary eclipse of the same target in an overlapping wavelength
regime have been published by \cite{Bell2017}. The $V$ band
corresponds to their two reddest wavelength bins, which resulted in a
depth of \mbox{0.06 $\pm$ 0.1 ppt} when averaged. This value is in
good agreement to the result of our second event, however, our first
and third event results deviate by 6\,$\sigma$ and 2.5\,$\sigma$,
respectively.

Potentially, systematic noise might mimic the deep ingress and egress
features of our first eclipse event. However, we find no abnormal
behaviour in any instrumental parameters at this time. Also the very
close match of the ingress/egress timings to the predicted times makes
an instrumental or systematic origin of these features less
likely. The literature offers additional hints for a time-variability
of the eclipse depth of \mbox{WASP-12b}. Observations obtained in the
Sloan $z'$ band at 0.9~$\mu$m by \cite{Lopez2010} and
\cite{Fohring2013} disagreed at the 2.4\,$\sigma$ level. Very
recently, \cite{Hooton2019} measured two eclipses of WASP-12b in the
Sloan i' band that differed by $\sim$\,3\,$\sigma$. Also for another
very hot Jupiter, \mbox{WASP-19b}, discrepant results have been
published by \cite{Burton2012} and \cite{Lendl2013} that might
potentially be explained by variability in time.

We assume that the origin of the different eclipse depths of this work
is a variability in time, and want to briefly discuss potential
sources. The depth of a secondary eclipse is the sum of light
reflected by the planet towards the observer and thermally emitted
light from the planet. We use Equation 4 and 5 of \cite{Alonso2018}
for a rough estimation of the potential time variability of these two
components. The thermal radiation of planet and host star is
approximated by black body radiation. For the involved stellar and
planetary parameters, we adopt the values of \cite{Collins2017}. If we
keep the thermal component fixed to a planetary black body radiation
of 2900~K \citep{Stevenson2014}, our secondary eclipse values can be
explained by a range of the geometric albedo from zero to
$\sim$\,0.75. Reflective clouds formed by large-sized condensates
might cause a geometric albedo of up to 0.4 \citep{Sudarsky2000}, and
we are not aware of any physical mechanisms causing a higher albedo in
the dayside atmosphere of \mbox{WASP-12b}, thus a geometric albedo of
0.75 appears unlikely.

If the variability of the eclipse depth is purely explained by an
increase in thermal radiation, a planetary temperature of about 4000~K
is needed. However, such an increase in planetary temperature by
1000~K would result in even more pronounced depth variations at NIR
wavelengths. The rough agreement of all existent NIR secondary eclipse
data of \mbox{WASP-12b}, taken at different epochs, to a black body
temperature of about 3000~K \citep[e.g.][]{Cowan2012,Stevenson2014},
makes a significant time variability of the photospheric temperature
unlikely.

Could both components act together to produce the large eclipse depth
of our first measurement? If we assume the temporary formation of a
reflective cloud deck causing a geometric albedo of 0.4, our simple
estimation results in an additional need of an increased black body
temperature to $\sim$\,3600~K. However, a spontaneous cloud formation
would intuitively be associated with a decrease in temperature instead
of an increase, therefore it is unclear under which circumstances the
reflective light component and the thermal light component can be
enhanced at the same time. Additionally, we note that in the dayside
of WASP-12b no cloud formation is expected because the very high
temperatures do not allow for particle condensation
\citep{Wakeford2017}.

The WASP-12 system is known to contain material eroded and blown off
from the planetary atmosphere by the extreme stellar irradiation
\citep{Fossati2013}. We can only speculate if potentially the variable
eclipse depth is not caused by the planetary dayside atmosphere, but
by an in-homogeneous flow of escaping material. This material might
form temporary clumps near the planet, which scatter a fraction of the
star light towards the observer. In this scenario, the deep secondary
eclipse would rather be caused by an occultation of the escaping
material than by the occultation of the planet.

\begin{table}[ht!]
    \centering
    \caption{\label{tab:WASP-12} Transit parameters (TPs) for
      WASP-12b, following the same description as
      Table~\ref{tab:TrES-3}.}
    \begin{tabular}{l c }
    \hline\hline
    TPs                              &  \cite{Maciejewski2018}       \\
    \hline
    P (days)                         & 1.09142172 $\pm$ 0.0000000015   \\
    T$_0$* (BJD$_\mathrm{TDB}$)      & 508.97694 $\pm$ 0.000005 \\
    dP/dE (days/epoch$^2$)           & (-9.67 $\pm$ 0.73) $\times$10$^{-10}$ \\
    a/R$_s$                          & 3.026 $\pm$ 0.02 \\
    i ($^{\circ}$)                   & 82.87 $\pm$ 0.4 \\ 
    R$_p$/R$_s$                      & 0.1175 $\pm$ 0.0003 \\
    u$_1$, u$_2$                     & 0, 0\\
    \hline
    \end{tabular}
\end{table}

\begin{table}[ht!]
    \centering
    \caption{\label{tab:WASP-12_SEDs} {\it From left to right:} Date
      of observation in dd.mm.yyyy, the secondary eclipse depths of
      WASP-12b, $\delta$, in ppt, their corresponding $\beta$ values,
      and the photometric precision of the data,
      $\sigma_\mathrm{phot}$, in ppt.}
    \begin{tabular}{l c c c}
    \hline\hline
    Date         &  $\delta$       &   $\beta$    &   $\sigma_\mathrm{phot}$  \\
    (dd.mm.yyyy) &  (ppt)          &              &   (ppt)                   \\
    \hline
    2019.01.01   & 1.16 $\pm$ 0.17 &   1.09       &    0.52                   \\
    2019.01.12   &-0.03 $\pm$ 0.17 &   1.60       &    0.59                   \\  
    2019.01.24   & 0.28 $\pm$ 0.10 &   1.10       &    0.57                   \\
    \hline
    \end{tabular}
\end{table}

\section{Conclusions}
\label{sec:conclusions}

In this work we present the characterization of two brand new
engineered diffusers (EDs) mounted at the 2.5 meter Nordic Optical
Telescope. The diffusers have two slightly different diffusing angles,
spreading the light of the stars differently over the CCD. They can be
placed in two different filter wheels and can work simultaneously to
several photometric filters placed on a third wheel. To characterize
the stability and reliability of the EDs we observed two photometric
standard stars. We found the throughput of the diffusers to be
superior to 90\% in the visible and for redder wavelengths, dropping
down to $\sim$75\% in the ultraviolet, as expected from their
construction and design. Both the core and the wings of the diffused
stellar images are quite stable as observations evolve, without
significant changes that correlate with seeing or
airmass. Furthermore, we observed three exoplanetary systems, namely
\mbox{TrES-3b}, \mbox{CoRoT-1b} and \mbox{WASP-12b}. While the first
two were observed during one primary transit each, the latter was
followed-up during three secondary eclipse events. Since the EDs can
be combined to regular photometric filters, our science frames were
taken using the Bessel $B$, $V$, and $R$ filters. The achieved
photometric precision of 0.5 ppt for a 25 second exposure in
\mbox{TrES-3b}, 0.65 ppt for 162 seconds for \mbox{CoRoT-1b}, and
between 0.52 and 0.59 ppt for less than one minute cadence for
\mbox{WASP-12b}, clearly shows the power of engineered diffusers to
achieve high precision photometric data. In addition to this, we
present a new planet-to-star radii ratio for \mbox{CoRoT-1b}, to our
knowledge the bluest data point taken to date, and we observe a
significant variability in the eclipse depth of \mbox{WASP-12b}, as
pointed out by other authors before in other optical
wavelengths. Within our collected data we observe a deviation within
eclipse depths of about 5\,$\sigma$. From previous observations in a
similar wavelength range, our first and third eclipse depths deviate
by 6\,$\sigma$ and 2.5\,$\sigma$. The reasons causing this variability
might be purely speculative.

\begin{acknowledgements}

CvE acknowledges funding for the Stellar Astrophysics Centre, provided
by The Danish National Research Foundation (Grant DNRF106), and
support from the European Social Fund via the Lithuanian Science
Council grant No. 09.3.3-LMT-K-712-01-0103. This work made use of
PyAstronomy\footnote{\url{https://github.com/sczesla/PyAstronomy}}. The
data presented here were obtained with ALFOSC, which is provided by
the Instituto de Astrofisica de Andalucia (IAA) under a joint
agreement with the University of Copenhagen and NOTSA. GKS
acknowledges support by NASA Headquarters under the NASA Earth and
Space Science Fellowship Program - Grant NNX16AO28H. JF and SD
acknowledge funding from the German Research Foundation (DFG) through
grant DR 281/30-1 and DR 281/32-1.

\end{acknowledgements}

\bibliographystyle{aa}
\bibliography{main}

\end{document}